\documentclass{article}
\usepackage[utf8]{inputenc}
\usepackage{graphicx}
\usepackage{amssymb}
\usepackage{amsmath}
\usepackage{cmll}
\usepackage{booktabs}
\usepackage{ dsfont }
\usepackage{cite}  

\usepackage[utf8]{inputenc} 
\usepackage[T1]{fontenc}    
\usepackage{hyperref}       
\usepackage{url}            
\usepackage{booktabs}       
\usepackage{amsfonts}       
\usepackage{nicefrac}       
\usepackage{microtype}      
\usepackage{lipsum}		
\usepackage{graphicx}
\usepackage{doi}
\usepackage{indentfirst}
\usepackage{multirow}
\usepackage{float}

\title{Model based reinforcement learning for stock trading optimization}
\author{Huifang Huang, Ting Gao, Luxuan Yang, Yi Gui, Jin Guo, Peng Zhang}
\date{September 2021}

\begin{document}

\maketitle

\begin{abstract}
Reinforcement learning (RL) is gaining attention by more and more researchers in quantitative finance as the agent-environment interaction framework is aligned with decision making process in many business problems. Most of the current financial applications using RL algorithms are based on model-free method, which still faces stability and adaptivity challenges. As lots of cutting-edge model-based reinforcement learning (MBRL) algorithms mature in applications such as video games or robotics, we design a new approach that 
leverages resistance and support (RS) level as regularization terms for action in MBRL, to improve the algorithm's efficiency and stability. From the experiment results, we can see RS level, as a market timing technique, enhances the performance of pure MBRL models in terms of various measurements and obtains better profit gain with less riskiness. Besides, our proposed method even resists big drop (less maximum drawdown) during COVID-19 pandemic period when the financial market got unpredictable crisis. Explanations on why control of resistance and support level can boost MBRL is also investigated through numerical experiments, such as loss of actor-critic network and prediction error of the transition dynamical model. It shows that RS indicators indeed help the MBRL algorithms to converge faster at early stage and obtain smaller critic loss as training episodes increase.
	
\end{abstract}

\section{Introduction}
As the development of modern deep learning techniques, more and more cutting-edge neural network structures have been widely used in various research and application fields. These machine learning frameworks in return inspire researchers from different background to design better neural network architectures to improve the model performance and solve more complex scenarios. One of the promising applications, quantitative finance, has recently been taking advantage of AI techniques to make lots of innovative algorithmic trading strategies for financial investment and portfolio optimization.

Among all the machine learning frameworks, reinforcement learning (RL) has an unique advantage that the interactive training process is aligned with human decision making. During the past years, two basic optimization ways, Q-learning and Actor-Critic have been developed to solve many RL problems, as well as the combination of the two. Take financial applications for example, Pastore et al.~\cite{2016Modelling} analyze data on 46 players from online games in financial markets and test whether the Q-Learning could capture these players'  behaviours based on a riskiness measure. Their results indicate that not all players are short-sighted, which contradicts with the naive-investor hypothesis. Besides, Li et al.~\cite{2019An} study the benefits of the three different classical deep RL models DQN~\cite{mnih2013playing}, Double DQN \cite{van2016deep} and Dueling DQN~\cite{wang2016dueling} by predicting the stock price and conclude that DQN has the best performance. In addition, some researchers simulate and compare the improved deep RL method with the Adaboost algorithm and propose a hybrid solution. Interestingly, Lee et al. (2019)~\cite{2019Global} apply CNN to a deep Q-network which takes stock prices and volumes chart images as input to predict global stock markets. On the policy gradient side, Kang et al. (2018)~\cite{2018An} utilize the state-of-art Asynchronous Advantage Actor-Critic algorithm (A3C~\cite{mnih2016asynchronous}) to solve the portfolio management problem and design a standalone deep RL model. Moreover, Li et al.~\cite{2019Optimistic} propose an adaptive deep deterministic RL method (Adaptive DDPG) for some portfolio allocation task, which incorporates optimistic or pessimistic deep RL algorithm based on the influence of prediction error. Through analyzing data from Dow Jones 30 component stocks, the trading strategy outperforms traditional DDPG method~\cite{lillicrap2015continuous}. Sutta~\cite{2019Robust} compares the performance of an AI agent to the results of the buy-and-hold strategy and the expert trader by testing 15 years' forex data market with a paired t-Test. The findings show that AI can beat the buy $\&$ hold strategy and commodity trading advisor in FOREX for both EURUSD and USDJPY currency pairs.

However, there are still complex and challenging issues with reinforcement learning for financial environment. First of all, financial marketing data, in most scenarios appear to have complex non-linearity, even complex chaotic behaviors and uncertainties including non-Gaussian noise, which causes finance time series data has distribution shift~\cite{cai2020distributed} over time. Besides, hidden interactions among different agencies can be unpredictably difficult to understand. As it is shown from lots of stock marketing analysis, agencies with large capital investment can sometimes cause dramatic price fluctuation, leading to marketing panic and instability among the public. Therefore, scholars are actively seeking intrinsic mechanics to solve the above problems. On one hand, various data denoise methods are utilized to preprocess the financial time series data. For example, Bao W, et al. (2017)~\cite{2017A} propose a new model for stock prediction, wavelet transforms for decomposing and eliminating noise of the stock price time series, which helps stacked auto-encoders (SAEs) for extracted high-level deep features and long-short term memory (LSTM) to better forecast the stock price and improve predictive accuracy and profitability performance. On the other hand, researchers also create many kinds of indexes to optimize portfolio gain. In short, all these make it hard for us to build an reinforcement learning system which directly interacts with complex real marketing environment, and this inspires us to investigate model-based RL for algorithmic trading strategy optimization.

Most of the existing literature is based on model-free reinforcement learning (MFRL) method. However, these algorithms are quite expensive to train, considering high risk of investment loss at early stage in real financial market environment, even using simulated domains (Mnih et al., 2015~\cite{mnih2015human}, Lillicrap et al., 2015~\cite{lillicrap2015continuous}, Schulman et al., 2017~\cite{schulman2017proximal}). A promising direction for improving sample efficiency is to explore model-based reinforcement learning (MBRL) methods ~\cite{2019Model}. DynaQ~\cite{peng2018deep}, PETS~\cite{chua2018deep}, MBPO~\cite{janner2019trust} are powerful model-based reinforcement algorithms in Gym environment. Model-based RL algorithms can attain excellent sample efficiency, but often lag behind the best model-free algorithms in terms of asymptotic performance. Chua, Calandra et al. (2019) ~\cite{chua2018deep} study how to bridge this gap by employing uncertainty-aware dynamical models. They propose a novel algorithm called PETS, which approaches the asymptotic performance of several benchmark model-free algorithms, while requiring significantly fewer samples. Janner et al. (2019)~\cite{janner2019trust} propose model-based Policy Optimization (MBPO) algorithm to boost PETS through monotonic improvement at each step and get state-of-the-art performance. All these excellent work inspires us to apply model-based RL algorithms into the complex financial marketing environment, and to the best of our knowledge, rare literature has been found on utilizing MBRL into quantitative financial modeling.

Quantitative finance is an interdisciplinary subject that involves many researchers from various fields with different background knowledge, and lots of promising algorithmic trading strategies in real-world problems utilize some financial technique analysis. For example, to find the best market timing for trading, resistance level (the price at which the trader believes that the seller’s power begins to overtake the buyer’s) and support level (the price at which traders believe that the buyer’s power begins to overtake the seller’s) are often considered as selling and buying market timing indicators respectively. Therefore, we leverage the timing selection power of Resistance Support Relative Strength (RSRS) and apply them into the classical MBRL algorithms, aiming to find better optimized policies that can improve the MBRL algorithms to be more stable, adaptive and efficient.


Overall, this paper makes the following major contributions: 
\begin{itemize}
    \item\textbf{Marriage of RSRS and MBRL}. 
    We design a financial environment for classic model-based reinforcement learning (MBRL) algorithms and equip  Resistance Support Relative Strength (RSRS) into the RL framework to improve model stability and adaptivity. 
   
    \item\textbf{Business Performance}.
    The Resistance Support Relative Strength (RSRS), as a powerful timing selection technique, when embedded into a couple of MBRL algorithms, help pure MBRL strategy get high improvement in seven different measurements, from both risk management and portfolio optimization point of view.
    
    
    \item\textbf{Model Performance}.
    To better explain the performance discrepancies between MBRL with and without RSRS as regularization for the actions, we also compare the loss of actor-critic network among all the algorithms. From the plots of critic loss, we can see that RSRS indeed help the selected MBRL algorithms to converge faster at early stage and obtain lower error when the models are convergent, therefore improves the model efficiency.

    \item\textbf{Performance of Transition Dynamics}.
    There is some close connection between SAC for policy optimization under transition dynamic modeling and stochastic optimal control with stochastic differential equations as state constraints. We also check the prediction error of the transition dynamical model in terms of different coordinates projections, considering our state space is complex and high dimensional.
  
\end{itemize} 


We structure our paper as the following sessions: First, in Section 2.1, we introduce some background knowledge of model-based RL, and then explain our financial environment setup under model-based RL framework in Section 2.2 and 2.3. We also interpret the transition dynamic from a dynamical system point of view and describe the Gaussian Process as some stochastic differential equation~(SDE), and correspondingly, the policy optimization of SAC is connected with stochastic optimal control under the constraint of a SDE in Section 2.4. Next, we explain the RSRS algorithm in detail in Section 3, and present the pseudo code of our proposed model. Moreover, all the experiment results are shown in Section 4, as well as detailed explanations on how and why our proposed algorithm is better than pure MBRL without RSRS. Finally, we summarize our conclusions and future research directions in Section 5.


\section{Framework}

\subsection{Background: Model-based RL}

Reinforcement learning aims to learn an optimal policy so as to maximize the expectation of cumulative rewards in the process of interacting with the environment. It mainly includes model-free and model-based methods in terms of whether the agent interacts with purely real environment.

Model-free methods can be used to handle many complex tasks with the best progressive performance. But usually this requires a lot of interaction with the environment and has relatively high requirements of computing power. However, model-based methods focus on building the environment model which we called transition dynamic model, with high sampling efficiency. One of the goals usually lies in improving the accuracy of estimations on state transition probability. Moreover, when the learned transition model is close to the real environment, the optimal strategy can be found directly through some planning algorithms.The traditional planning pipeline includes two parts: path search and trajectory optimization.

To be more specific, for path search, when the action space is continuous, CEM (Cross-Entropy Method) is usually used. It is a Monte Carlo method mainly used for optimization and importance sampling. When the action space is discrete, the MCTS (Monte Carlo Tree Search) is considered when the search space is huge. Trajectory optimization refers to the problem of solving the objective function with constraints, that is, the optimal control problem. Trajectory optimization generally includes shooting and collocation methods. Shooting methods are more suitable for simple control and problems with no path restrictions while collocation methods are more suitable for complex control and problems with path restrictions.

In PETS, parametric models fitted by a neural network, and propose the algorithm by combining the PE (Probabilistic Ensembles) model and the TS (Trajectory Sampling) planning method. With cross entropy method (CEM), they point out a better direction for the action, then sample based on it. Due to a certain deviation between the established environmental model and the real model, Janner and Fu et al. (2019)~\cite{janner2019trust} proposed a novel model-based algorithm MBPO to improve PETs. After formulating and analyzing a model-based RL with monotonic improvement at each step, they study the model usage in policy optimization both theoretically and empirically. Then they demonstrate that a simple procedure of using short rollout planning from real data and SAC (soft actor-critic)~\cite{haarnoja2018soft} for optimization.



 In complex and noisy environments, transition model error can be large. Learning a good policy requires an accurate model, while getting an accurate model requires a lot of interaction with the real environment. 
 Pan and He (2020) proposed M2AC (Masked Model based Actor-Critic)~\cite{pan2020trust}, which reduces the influence of model error through masking mechanism, and alleviates the overfitting problems of MBRL in the case of small data size and large noise in real environment. It is theoretically proved that the gap between the real return and the masked ensemble model rollout value can be bounded when the individual model being used is with a small error.


\subsection{Space and Reward Definition}
By considering our agency (investor) as the intelligent agent and financial market as the corresponding environment, we model the stock trading problem as a Markov Decision Process (MDP). 


$\bullet$ \textbf{State space} $\mathcal{S} = [B, P, W, I]$:
State space $\mathcal{S}$ is the market information collected. $\forall s_t \in \mathcal{S}$ is the state of the agent at time $t$, which includes four parts information: account balance for agency $B_t$, current stock price $P_t$, cumulative holding $W_t$, technical indicators $I_t$. Here the technical indicators $I_t$ is consisted of seven common technology factors: MACD~\cite{appel2003become}, SMA30, SMA60, BOLL~\cite{bollinger1992using}, RSI~\cite{ctuaran2011relative}, CCI~\cite{lai2020bidirectional}, ADX. Table 1 describes the notations of state $s_t$ in detail.

We assume the experiment consist of D stocks.
\begin{table}
\caption{Notations of State }
\label{table_1}
\centering
\begin{tabular}{l|lll} 
\hline
Notation & Defination \\ 
\hline
$B_t$ & \begin{tabular}[c]{@{}l@{}} Account balance at time t;  $B_t \in \mathbb{R}_+$ \end{tabular} \\
\hline
$P_t$ & \begin{tabular}[c]{@{}l@{}}  daily closing price of each stock; $P_t \in \mathbb{R}_+^{D}$  \end{tabular} \\
\hline
$W_t$ & \begin{tabular}[c]{@{}l@{}} cumulative holding shares of each stock; $W_t \in \mathbb{Z}_+^{D}$  \end{tabular} \\
\hline
MACD & \begin{tabular}[c]{@{}l@{}} Moving Average Convergence Divergence: \\a momentum indicator displays trend \end{tabular} \\
\hline
SMA30 & \begin{tabular}[c]{@{}l@{}} 30 day Simple Moving Average:\\ 30 day closing price equal weighted average   \end{tabular} \\
\hline
SMA60 & \begin{tabular}[c]{@{}l@{}} 60 day Simple Moving Average: \\60 day closing price equal weighted average   \end{tabular} \\
\hline
BOLL & \begin{tabular}[c]{@{}l@{}} Bollinger Bands:\\ judges the medium and long-term movement trend   \end{tabular} \\
\hline
RSI & \begin{tabular}[c]{@{}l@{}} Relative Strength Index: \\identifies inflection points of a trend  \end{tabular} \\
\hline
CCI & \begin{tabular}[c]{@{}l@{}} Commodity channel index: \\helps to find the degree of deviation of the price  \end{tabular} \\ 
\hline
ADX & \begin{tabular}[c]{@{}l@{}}  Average Directional Index:\\ determines the strength of a trend  \end{tabular} \\
\hline
\end{tabular}
\end{table}

$\bullet$ \textbf{Action Space} $\mathcal{A}$: Action space $\mathcal{A}$ is a set of available operations during the transaction on all D stocks. $\forall a_t \in \mathcal{A}$ is the action taken by the agent at time $t$, assumed to be finite dimensional and continuous. Here $a_t$ is a D-dimensional vector, where the $i_{th}$ dimension represents the action performed on the $i_{th}$ stock.
 
Noting $W^i_t$ as the cumulative shares of $i_{th}$ stock in time $t$, the available actions of the stock include buying, holding and selling. We assume the maximum trading volume for a single stock at one step is 100 shares. The details as follows:\\
• Buying: $a^i_t = + h$, $h$ shares can be bought and it leads to $W^i_{t+1} = W^i_t + h$.\\
• Selling: $a^i_t = - h$, $h$ shares can be sold from the current holdings. In this case, $W^i_{t+1} = W^i_t - h$.\\
• Holding: $a_t = 0$, that means no change in $W^i_t$.\\
where $h \in [0,100]$ is a positive integer.

$\bullet$ \textbf{Reward} $r_t$: The reward function is a mapping $\mathcal{R}: \mathcal{S} \times \mathcal{A} \rightarrow \mathbb{R}$. As is shown in later equation (3), the direct reward $r_t$ of taking the action $a_t$ in state $s_t$ is defined as the percentage of asset amount change:

1)The asset amount of the agent is the sum of the remaining investment funds and the current value of stocks holding. We note the asset amount of the agent in time $t$ is $Asset_t$, and the calculation formula as equation:
\begin{equation}
    Asset_t = (B_t + P_t^T \cdot W_t)
         = B_0 - \sum_{\tau = 0}^t P_{\tau}^T \cdot a_{\tau} + P_{t+1}^T \sum_{\tau = 0}^t a_{\tau}
\end{equation}

2) Consider the transition cost, we notes the cost as $C_t$:
\begin{equation}
    C_t = P_t^T \cdot|a_t|\cdot cost_{percentage}
\end{equation}
the absolute value of $a_t$ here means taking the absolute value of each component of a without changing the dimension of the vector $a_t$.

3) The calculation formula of direct reward $r_t$ is:
\begin{equation}
    r_t = \frac{Asset_{t+1} - Asset_t - C_t}{Asset_{t}} \times 100
\end{equation}
where $B_t,P_t$ is shown as Table 1; $a_t$ is the action given by agent at time $t$;$P_t^T \cdot W_t$ is the inner product of vectors.

For RL problem, the agent’s objective is to maximize its expected reward. The reward can assess the quality of each action. What's more, the agent optimizes the strategy under the guidance of the reward. Noting the cumulative reward as $R_t$ in time t:
\begin{equation}
    R_t = \Sigma_{\tau=1}^t  r_{\tau}
\end{equation}


\subsection{Transition Dynamic}
One of the classification criteria of RL algorithms is whether dynamic model of the environment exists. Model-based RL is appealing because the dynamic model is reward-independent and it can easily benefit from all of the advances in deep supervised learning to utilize high-capacity models~\cite{chua2018deep}. However, the disadvantage of this method is that in many learning tasks, the agent is often hard to obtain an accurate model of the real environment, leading to big accumulated error when interacting with virtual environment. PETS, MBPO and M2AC are robust algorithm to take a step toward narrowing the gap between model-based and model-free RL methods. Therefore, we apply these models to the financial market for stock trading optimization. We start with the definition of transition probability in the following.



$\bullet$ \textbf{Transition Probability} $\mathcal{P}$: The change of market conditions is abstracted as the state transition function. $\mathcal{P}: S \times A \times S \rightarrow [0,1]$ is a function of probabilities of state transitions. $$P_{s,s'}^a = \mathcal{P}(s_{t+1} = s' | s_t = s, a_t = a)$$.

In PETS, MBPO and M2AC, ensemble Gaussian Process is used to predict state transition dynamics. From stochastic analysis, we know that the state $s_t$ satisfies the following stochastic differential equation (SDE):
\begin{equation}
    dX_t = b(X_t,u_t)d_t + \sigma(X_t,u_t)dw_t\label{SDE_control}
\end{equation}
where the initial condition $X_0 = x \in \Omega$, and $a_t \subset \mathbb{R}^{da}$ is an $\mathbb{F}_t$-adapted control field and $w_t$ is a $d_w$-dimensional $\mathbb{F}_t$-standard Brownian Motion.\\

$\bullet$ \textbf{Policy} $\pi$: Policy is a mapping characterized by the policy $\pi :\mathcal{S} \rightarrow \mathcal{A}$. $\forall t \in {1, . . . , T}$. It is an action sequence given by agent, the objective of policy $\pi$ given by agent is to maximize its finally expected portfolio value.

The agent in state $s_t \in \mathcal{S}$ takes an action $a_t \in \mathcal{A}$ follows policy $\pi$, receives the reward $r_t = \mathcal{R}(s_t, a_t)$ and transits to the next state $s_{t+1}$ according to the transition probability $\mathcal{P}$. The agent interacts with the environment to collect the trajectory, and then updates the strategy. The core optimization problem in RL is to find the optimal strategy $\pi^*$ that optimize total accumulated reward. 


\subsection{Policy Optimization}
 The actor-critic bridges the gap between value function approximation methods and policy gradient methods for RL. This approach has proven to be able to learn and adapt to large and complex environments, and has been used to play popular video games, such as Doom~\cite{2017training}. Thus, the actor-critic approach is applying for trading with a large stock portfolio~\cite{xiong2018practical}. The Proximal Policy Optimization (PPO) Algorithm(Schulman et al., 2017~\cite{schulman2017proximal}) is an actor-critic off-policy model-free RL algorithm  for discrete control and continuous control. However, it faces serious sample inefficiency and requires a huge amount of sampling to learn, which is unacceptable for real stock trading market training. Another typical example of actor-critic algorithms is Deep deterministic policy gradient (DDPG) algorithm (Lillicrap et al., 2015~\cite{lillicrap2015continuous}). Researchers explore the potential of RL for stock trading with DDPG algorithm and achieve some good results~\cite{emami2016deep,azhikodan2019stock}. While it is very sensitive to various hyper-parameters during actual training, so the excellent performance of DDPG on various benchmarks is actually crafted and it is difficult to solve a lot of specific problem.

Tuomas Haarnoja proposed SAC Algorithm~\cite{haarnoja2018soft}, an off-policy method, to solve some issues of sampling inefficiency in PPO, and sensitivity of hyper-parameters in DDPG. The biggest difference in SAC is its objective of entropy maximization, which is a great benefit for maintaining stochasticity in policy selection, and hence can help agent to search state space more completely so as to avoid local optimal and sensitivity.

For random variable $x$ with probability density $p$, the entropy $H$ of $x$ can be defined as:
\begin{equation}
    H(p) = \underset{x\thicksim p}{\mathbb{E}}[-log p(x)]
\end{equation}
Consider entropy regularized RL, the reward will be added with an extra term which is the entropy. At the same time, the influence of current action on future rewards will be weakened with time going by, so we use discounted reward instead of the directed reward. The soft value function (soft Q function) becomes:
\begin{equation}
\begin{aligned}
    Q_{soft}^\pi(s, a)= & \underset{(s_t,a_t)\thicksim \rho_\pi}{\mathbb{E}} [\Sigma_{t = 0}^\infty \gamma^tr(s_t, a_t) + \\
    & \alpha \cdot \Sigma_{t=1}^\infty \gamma^t H(\pi(\cdot|s_t)) | s_0=s,a_0=a]
\end{aligned}
\end{equation}
where $\rho_\pi$ represents the distribution of state-action pair of the agent under policy $\pi$.
Correspondingly, soft V function becomes:
\begin{equation}
    V_{soft}^\pi(s) = \underset{(s_t,a_t)\thicksim \rho_\pi}{\mathbb{E}} [\Sigma_{t = 0}^\infty \gamma^t(r(s_t, a_t) + \alpha H(\pi(\cdot|s_t)))|s_0=s]
\end{equation}
With these definitions, $V^\pi$ and $Q^\pi$ and connected by
\begin{equation}
    V^\pi(s) = \underset{a\thicksim \pi}{\mathbb{E}}[Q^\pi(s,a) + \alpha H(\pi(\cdot|s))]
\end{equation}
Thus, the Bellman equation can be written as:
\begin{equation}
    Q_{soft}^\pi(s, a) = \underset{s'\thicksim \mathcal{P}(s'|s,a)}{\mathbb{E}} [r(s,a) +\gamma(Q_{soft}^\pi(s',a')+\alpha H(\pi(\cdot|s')))]
\end{equation}
 There are several versions of SAC for the past years. Here we use the older version that learns a value function $V$ in addition to $Q$ function.
\begin{equation}
    Q_{soft}^\pi(s, a) = \underset{s'\thicksim \mathcal{P}(s'|s,a)}{\mathbb{E}} [r(s,a) +\gamma(V_{soft}^\pi(s')]
\end{equation}
Thus, iteration of value function is
\begin{equation}
    V_{soft}^\pi(s) = \underset{a\thicksim \pi}{\mathbb{E}} [Q_{soft}^\pi(s, a) - \alpha log\pi(a|s)]
\end{equation}
By clipped donate Q trick, the loss function of Q-network in SAC is:
\begin{equation}
    \mathcal{L}(\phi,D) = \underset{(s,a,r,s') \in D)}{\mathbb{E}}[Q_\phi(s,a) - y(r,s',d)]^2
\end{equation}
where $Q_\phi$ and $Q_{target}$ are neural networks and 
\begin{equation}
    y(r,s',d) = r + \gamma(1-d)(\underset{j = 1,2}{min} Q_{target,j}(s',\tilde{a}') - \alpha log \pi_\theta(\tilde{a}'|s'))
\end{equation}
where $\tilde{a}'\thicksim \pi_\theta(\cdot|s')$ in which a sample from $\pi_\theta(\cdot|s)$ is drawn by Gaussian distribution.\\

Now we formulate this problem as a stochastic optimal control problem and the stochastic optimal control problem can be written as:
\begin{equation}
    \mathcal{J}^u(x) = \mathbb{E}[\int_0^\infty f(X_s,u_s)e^{-\gamma s} ds |X_0^u = x]
\end{equation}
constrained on equation (\ref{SDE_control}).\\
Define
\begin{equation}
V(x) = \underset{u}{inf}\mathcal{J}^u(x) - \alpha log \pi(a|s)
\end{equation}
Then V satisfies the time-dependent HJB equation:
\begin{equation}
\underset{u}{inf}{\mathcal{L}^u V(x,u) + f(x,u) - \gamma V(x)} = 0
\end{equation}
where $\mathcal{L}^u V(x) = \frac{1}{2} Tr(\sigma \sigma^T Hess(V))(x,u) + b(x,u)^T \nabla V(x)$ is the generator of SDE (\ref{SDE_control}).

\section{Marriage of MBRL and RSRS}

In this article, we use the Resistance Support Relative Strength (RSRS) market timing indicator \footnote{Junwei Liu, Xiaoxiao Zhou. Market Timing Based on Resistance Support Relative Strength (RSRS). \textit{Everbright Securities Technical Timing Series Report Series I}, 2017.5.1}. It is a technical indicator to choose the best time for buying and selling by measuring the relative strength of support and resistance~\cite{lloyd2013successful}. With the definitions of support level and resistance levels in the financial field, we have:

$\bullet$ \textbf{Resistance level}: The price at which the trader believes that the seller's power begins to overtake the buyer's, making it difficult for the price to continue to rise or to pull back from the falling price.

$\bullet$ \textbf{Support level}: The price at which traders believe that the buyer's power begins to overtake the seller's, thereby stopping the decline or rebounding the rising price.

The RSRS indicator no longer regards resistance and support as a fixed value, but as a variable in that report. This indicator is an expected judgment of traders on the top and bottom of the current market state. 

Researchers replaced the specific price threshold of support and resistance with the relative strength of the daily high and low price changes. Let $\mu$ be the mean value of the historical RSRS slope, and $\sigma$ be the standard deviation of the historical slope. Denote the resistance level as $S_{buy}$, where $S_{buy}=\mu+\sigma$. Then denote the support level as $S_{sell}$, where $S_{sell}=\mu-\sigma$. When the RSRS slope is greater than $S_{buy}$, buy the whole position, and when it is less than $S_{sell}$, sell and close the position. In this way, the drawdown can be controlled and the profit performance of the strategy will be guaranteed.

\subsection{RSRS Indicator}
The RSRS indicator uses the magnitude of the high price change when the low price of a certain stock changes by 1 unit to measure the strength of support and resistance. That is, to regress the high price series and the low price series in a certain period. The slope $\beta$ obtained by the model is a specific indicator for measuring resistance support relative strength.

Taking the high price time series and low price time series of the previous N days of a certain stock. We calculate the RSRS slope of the day: 
\begin{equation}
    high  = \alpha + \beta \times low + \epsilon
\end{equation}
where $\epsilon \backsim \mathcal{N}(0,\sigma_\epsilon^2)$ 

\subsection{Right Standard Score}
For stocks in different periods of the market, the mean value of the slope will fluctuate significantly. Therefore, it is not appropriate to directly use the mean value of the slope as the timing index. It is a good choice to standardize the slope value.

1) Taking the RSRS slope time series of the previous M days of a certain stock. We calculate the standard score $RSRS_{std}$ of the RSRS slope of the day: 
\begin{equation}
    RSRS_{std} = \frac{RSRS - \mu_M}{\sigma_M}
\end{equation}
where $\mu_M$ is the average slope of the previous M days and $\sigma_M$ is the standard deviation of the previous M days.

2) In fact, when the slope is used to quantify the relative strength of resistance support, its quantitative effect is largely affected by the effect of the fitting itself. So further, we consider the fitting effect and weight the standard scores with the coefficient of determination (the R-square value in the regression model) to reduce the impact of the RSRS standard score, which has a large absolute value but poor fitting effect, on the strategy.
\begin{equation}
    RSRS_{cor} = RSRS_{std} \times R^2
\end{equation}
where $R^2$ is the coefficient of determination in the regression model of the day.

3) The RSRS indicator has excellent left-side forecasting ability by itself. So we use the significant right-biased standard score correction as the timing index in this paper.
\begin{equation}
    RSRS_{rightdev} = RSRS_{cor} \times RSRS
\end{equation}

4) According to the analysis before, the RSRS timing index is defined as:
\begin{equation}
    RSRS_{rightdev} = \frac{RSRS - \mu_M}{\sigma_M} \times R^2 \times RSRS
\end{equation}

\subsection{Proposed Model}
Based on RSRS indicator explained in the previous session, we know it's a relatively strong policy to resist deep drop when a stock has big plunge during some urgent time period. Therefore, we believe that combining this strategy to classical RL algorithms can help to get the optimal strategy with less volatility and more stability, meaning that we could get higher Sharpe ratio and lower maximum drawdown. Among all the recent state-of-the-art model based RL algorithms, we choose MBPO and M2AC as our illustrated backbone models, while we think other models such as BMPO~\cite{lai2020bidirectional} and MVE~\cite{feinberg2018model} may also have similar results. The pseudo code is put in table \ref{table_2} which uses MBPO as an example, and the case with M2AC is similar and easy to obtain. For the sake of illustration, we note MBPO algorithm with RSRS indicator as RSPO strategy and  M2AC algorithm with RSRS indicator as RSAC strategy later.
\begin{table*}[!t]
\renewcommand{\arraystretch}{1.3}
\caption{RSPO Framework}
\label{table_2}
\centering
\begin{tabular}{l}
\toprule
\textbf{Algorithm of RSPO : MBPO algorithm with RSRS indicator} \\
\midrule
Initialize policy $\pi_\theta$, dynamic model $\mathcal{P}_\phi$, environment dataset $\mathcal{D}_{env}$, model dataset $\mathcal{D}_{model}$ \\
Initialize time window $l$, considering window M, buy threshold $rs_{buy}$, sell threshold $rs_{sell}$ and maximum action $hmax$ \\
\bf{for E epochs do}\\
\quad Train dynamic model $\mathcal{P}_\phi$ on $\mathcal{D}_{env}$ \\
\quad \bf{for N steps do}\\
\quad \quad $\vec{y}\:=\:\alpha\:+\:\beta_i \cdot \vec{x}\:+\:\epsilon,\:(i\:=\:1,\:...,\:N )$, where $\vec{y}$ and $\vec{x}$ corresponds to the high price \\
\quad \quad and low price sequence of length ${l}$ respectively \\
\quad \quad \bf{for M steps do} \\
\quad \quad \quad $\vec{y}\:=\:\alpha + \beta_m \cdot \vec{x} + \epsilon,\:(m\:=\:i-M+1,\:...,\:i )$ \\
\quad \quad ${\beta_{std}}_i\:=\:\frac{\beta_i\:-\:\mathbb{E}(\beta_j)}{\sigma(\beta_j)},\: j\:=\:i\:-\:M\:+\:1,\:...,\:i$ \\
\quad \quad ${\beta_{mod}}_i\:=\:{\beta_{std}}_i\:\times\:R^2$, where $R^2$ is the coefficient of determination corresponds to $\beta_i$ \\
\quad \quad ${\beta_{rightdev}}_i\:=\:\beta_{mod}\:\times\:\beta_i$ \\
\quad \quad Get action $a_t$ using policy $\pi_\theta$ \\
\quad \quad \bf{if} ${\beta_{rightdev}}_i\:>\:rs_{buy}$,~\bf{then} $a_t\:=\:hmax$ \\ 
\quad \quad\bf{if} ${\beta_{rightdev}}_i\:<\:rs_{sell}$,~\bf{then} $a_t\:=\:-\:hmax$ \\
\quad \quad Take action $a_t$ in environment ; add to $\mathcal{D}_{env}$\\
 \quad \quad \bf{for L model rollouts do}\\
 \quad \quad \quad Sample $s_t$ randomly from $\mathcal{D}_{env}$\\
\quad \quad \quad Perform k-step model rollout starting from $s_t$ using policy $\pi_\theta$ ; add to $\mathcal{D}_{model}$\\
 \quad \quad \bf{for W gradient updates do}\\
\quad \quad \quad Update policy parameters on $\mathcal{D}_{model}$: $\theta\:\leftarrow\:\theta\:-\:\lambda_\pi\:\triangledown_\theta\:(\mathcal{J}_\pi(\theta, \mathcal{D}_{model}))$\\
\bottomrule
\end{tabular}
\end{table*}

\section{Experiment}
In this session, we present some experiment results and corresponding explanations, from which we can see how and why RSRS helps improving the performance of model-based classical reinforcement learning models in stock portfolio.



\subsection{Dataset}
All of our sample data are from Yahoo finance database \footnote{https://www.yahoo.com}. We conduct our experiments on 30 stocks selected by their turnover rate within 180 days before Jan. 01, 2009 in The Standard and Poor's 500 (S\&P 500) market. The selected stocks have relatively low turnover rate so as to avoid big noise or fluctuations in unstable financial markets with small market capitalization. Here, as a liquidity indicator, the "turnover rate" refers to the frequency of stock changing hands in the market within a certain period of time. In this way, we can better explore the applicability and generalizability of the proposed agent in this paper.

For each experiment, the time span of the dataset is historical daily price data from Jan. 1, 2009 to Jul. 3, 2021. Data from Jan. 1, 2009 to Jul. 3, 2016 (\textbf{1888 days}) is used as training set, and data from Jul. 4, 2016 to Jul. 3, 2018 (\textbf{504 days}) is used as validation set. The remaining data from Jul. 4, 2018  to Jul. 3, 2021 (\textbf{755 days}) is used as testing set. We train our agent on training data, then select model hyper-parameters by evaluating metrics such as annualized return, Sharpe ratio and maximum drawdown on validation data, and finally apply the selected model onto the testing data, which is shown in the following figures and tables. Notice that all the testing results are averaged from 10 random experiments. 

Here we setup the following hyperparameters for our experiments: initial balance $B_0 = 1e6$ dollars, time window $l = 10$, time period of mean and standard deviation for RSRS $M = 300$, two thresholds $rs_{buy} = 1.0$, $rs_{sell} = -0.4$, and maximum shares per transaction $hmax = 100$. 

\subsection{Portfolio Performance}

Five algorithms, PETS, MBPO, M2AC, RSPO, RSAC are used in our experiments. All these strategies on testing data are compared with the Market Index baselines: GSPC.

The evaluation metrics, we use the following measurements to evaluate proposed model performance. 
\begin{itemize}
    \item{Annualized Return: the geometric average amount of money earned by an investment strategy each year over a given time period.}
    \item{Cumulative Return: reflects the overall effect of trading strategy in a certain time range.}
    \item{Annualized Volatility: annualized standard deviation of portfolio return shows the robustness of the agent.}
    \item{Sharpe Ratio~\cite{sharpe1998sharpe}: the return earned per unit volatility which is a widely-used measure of the performance of an investment.}
    \item{Calmar Ratio: the annualized return earned per unit maximum drawdown in the period.}
    \item{Stability: Determines R-squared of a linear fit to the cumulative log returns.}
    \item{Maximum Drawdown: the maximum loss from a peak-to-trough decline of an investment, before a new peak is attained.}
\end{itemize}

\noindent\textbf{Overall performance with different measurements:}  In Table \ref{table_3}, we use the above seven indicators to evaluate the performance of different RL algorithms. It can be seen from the table that between July 2018 and July 2021, all the model-based reinforcement learning algorithms are better than the baseline method (GSPC), especially RSAC and RSPO, which are ranked as top 2 methods in terms of all the measurements. To be more specific, first, from profit point of view, RSAC has around 30\% annualized return and 120\% cumulative return, which are nearly twice as high as those of GSPC. Second, from the risk management point of view, RSAC and RSPO have relatively low annualized volatility, which is aligned with high stability ratio and low maximum drawdown, meaning that RSRS can help the agent to make more stable investment decision with less riskiness. And finally, considering both risk management and profit gain, Sharpe ratio and Calmar ratio show that RSAC and RSPO are significantly better than others, indicating combining RSRS with some classical reinforcement learning algorithms is a worthy direction to try.

\begin{table*}[t!]
\renewcommand{\arraystretch}{1.0}
\caption{Strategy Performance in S\&P 500. The specific performance of the six strategies on seven backtest indicators during the three-years backtest period from July 2018 to July 2021.}
\label{table_3}
\centering
\begin{tabular}{l|cccccc} 
\hline
                   & RSAC              & RSPO              & M2AC     & MBPO     & PETS     & GSPC      \\ 
\hline
\begin{tabular}[c]{@{}l@{}} Annualized\\ Return \end{tabular}   & \textbf{30.02\%}  & 28.23\%           & 23.43\%  & 22.78\%  & 20.21\%  & 16.75\%   \\
\hline
\begin{tabular}[c]{@{}l@{}}  Cumulative \\Return \end{tabular} & \textbf{119.56\%} & 110.62\%          & 87.87\%  & 84.96\%  & 73.60\%  & 59.04\%   \\
\hline
\begin{tabular}[c]{@{}l@{}} Annualized \\ Volatility \end{tabular} & \textbf{22.78\%}  & 23.05\%           & 23.58\%  & 24.91\%  & 24.15\%  & 23.05\%   \\
\hline
\begin{tabular}[c]{@{}l@{}} Sharpe \\ Ratio \end{tabular}      & \textbf{1.27}     & 1.2               & 1.01     & 0.95     & 0.88     & 0.79      \\
\hline
\begin{tabular}[c]{@{}l@{}} Calmar \\ Ratio \end{tabular}  & 105.14\%          & \textbf{106.86\%} & 72.57\%  & 64.58\%  & 61.11\%  & 49.37\%   \\
\hline
Stability          & 94.63\%           & \textbf{95.54\%}  & 89.08\%  & 88.96\%  & 90.03\%  & 72.82\%   \\
\hline
\begin{tabular}[c]{@{}l@{}} Maximum \\ Drawdown \end{tabular}   & -28.55\%          & \textbf{-26.41\%} & -32.28\% & -35.28\% & -33.07\% & -33.92\%  \\
\hline
\end{tabular}
\end{table*}

\noindent\textbf{Comparison of Annualized Return:} Table \ref{table_4} shows the year-end returns of the six strategies for each year. Analysis of the tabular data shows that between July 2019 and July 2020, due to the impact of COVID-19, the stock market sentiment was very sluggish. The benchmark GSPC has only a yield of 4.66\%, and M2AC, MBPO, PETS, as our control groups, all have the highest yield less than 10\%. However, our proposed method with RSRS equipped into M2AC and MBPO, still has maintained the annualized return close to be around 20\% and 28\%, showing that the RSRS indicator can effectively resist  plunge when the finance market gets collapse during some unpredictable crisis.

However, from Jul. 2020 to Jul. 2021, the year following the rapid recovery of the stock market, the yields of M2AC and MBPO increased sharply, outperforming RSAC and RSPO. We speculate the reason for this might come from the low asset amount of M2AC and MBPO in Jul. 2020. 

Based on the above analysis, we believe that M2AC and MBPO algorithms are comparatively sensitive to market sentiment, which shows that the RSRS indicator could enhance the algorithms robustness against the violent fluctuation of the finance market in a certain extent.

\begin{table}
\caption{Annualized return of the six strategies in each year from July 2018 to July 2021.}
\label{table_4}
\centering
\begin{tabular}{l|ccc} 
\hline
     & 2018.7.1-2019.7.1 & 2019.7.1-2020.7.1 & 2020.7.1-2021.7.1  \\ 
\hline
RSAC & \textbf{29.90\%}  & \textbf{19.53\%}  & \textbf{38.33\%}   \\
RSPO & \textbf{31.72\%}  & \textbf{27.88\%}  & 22.66\%            \\
M2AC & 23.22\%           & 5.94\%            & \textbf{42.25\%}   \\
MBPO & 22.46\%           & 8.01\%            & 36.94\%            \\
PETS & 15.42\%           & 9.31\%            & 34.13\%            \\
GSPC & 9.51\%            & 4.66\%            & 36.87\%            \\
\hline
\end{tabular}
\end{table}

\begin{figure}
    \centering
    \includegraphics[width=0.4\textwidth]{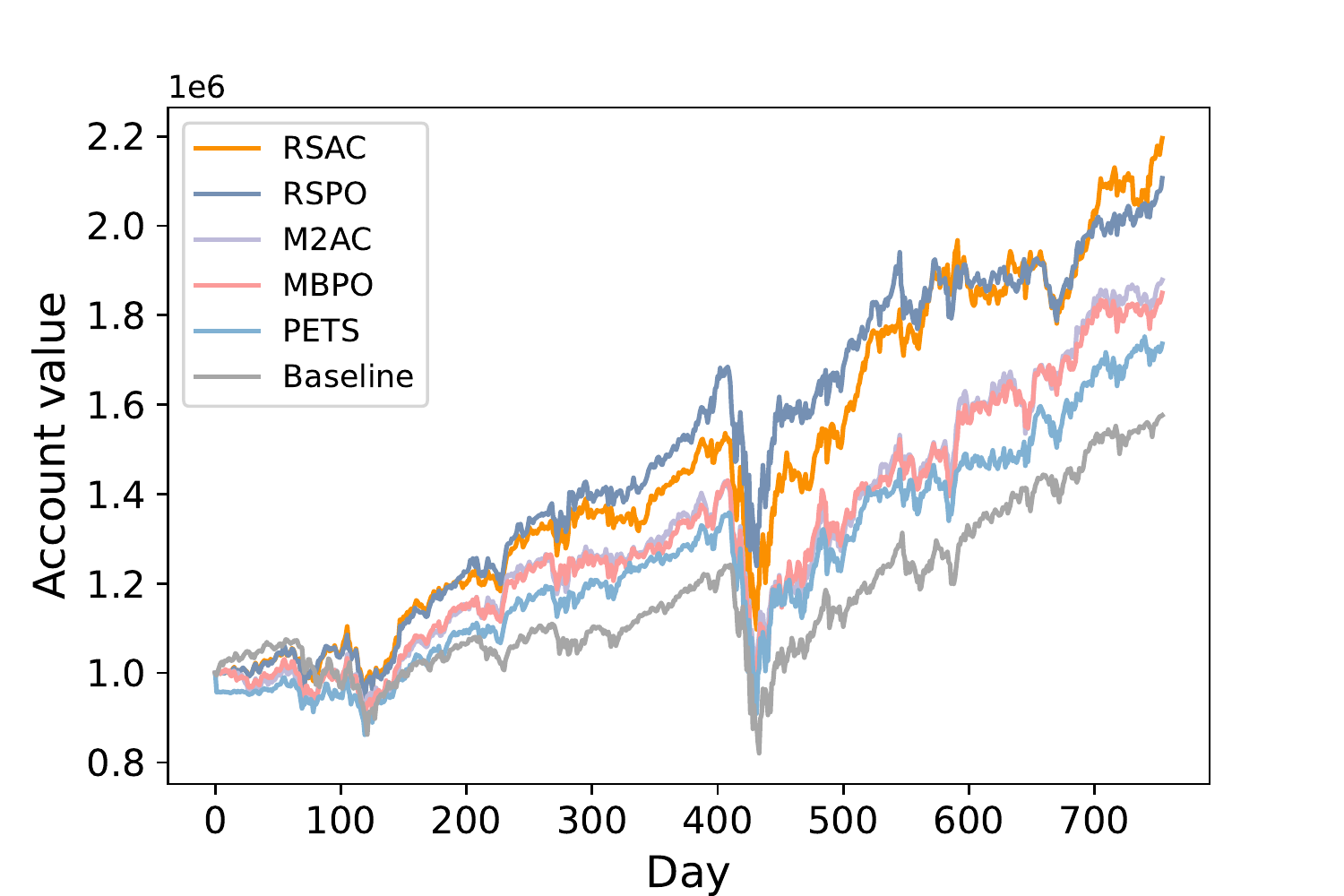}
    \caption{Returns. \,Time-yield curve of RSAC, RSPO and M2AC, MBPO, PETS, Baseline.}
    \label{returns}
\end{figure}

\noindent\textbf{Comparison of Daily Return:} For all 30 selected stocks, we show the portfolio of the five algorithms and the corresponding baseline in Figure~\ref{returns}. It illustrates the account value of six strategies in the experiment. Compared the performance of RSAC, RSPO, M2AC and MBPO, it can be seen that after adding the RSRS indicator, the income of RSAC (orange) and RSPO (dark blue) has increased significantly and is consistently higher than the cumulative income of PETS (light blue) and baseline (grey).

\subsection{Model Performance}

\begin{figure}
    \centering
    \includegraphics[width=0.4\textwidth]{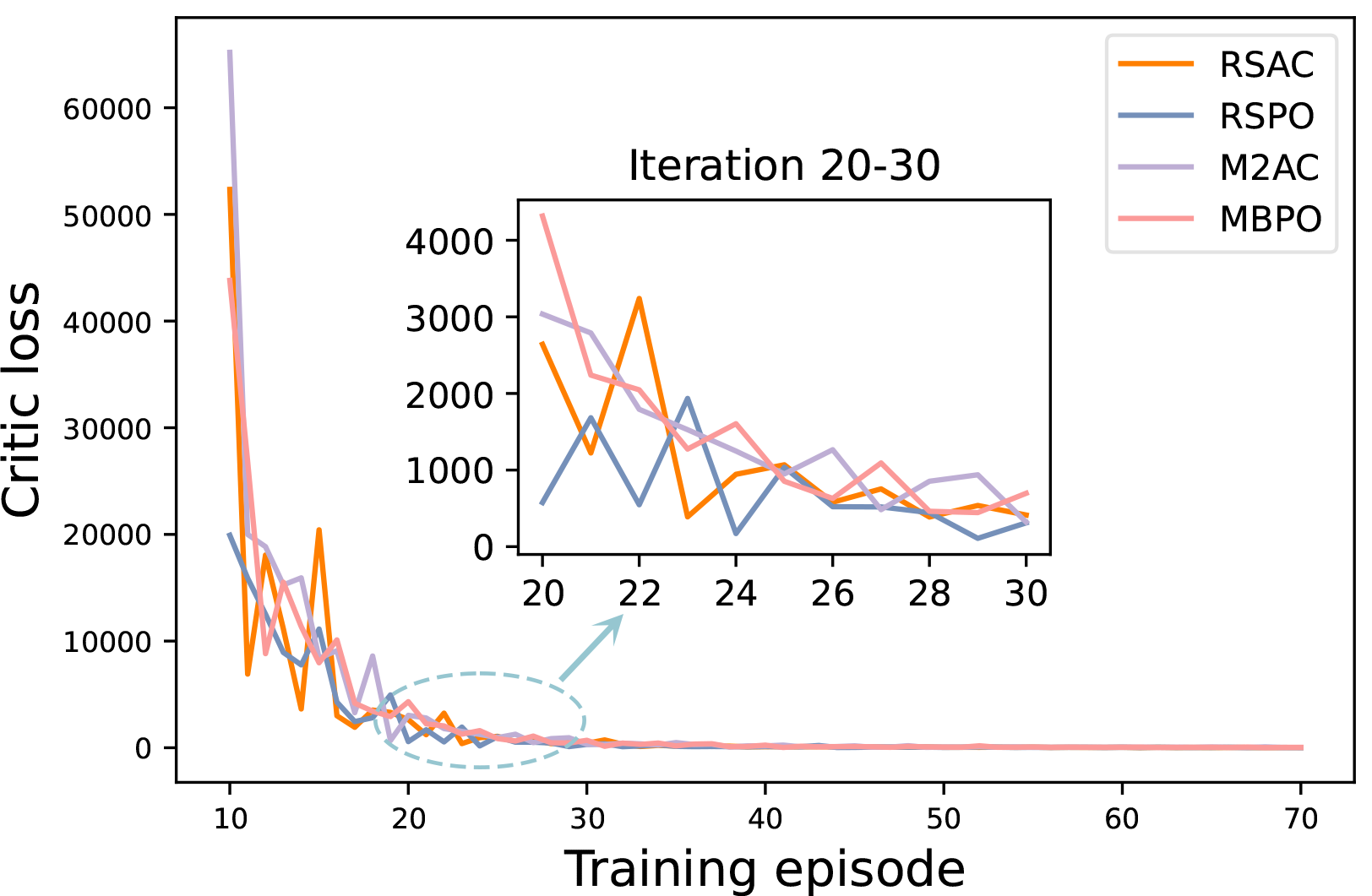}
    \caption{Critic loss over training steps.}
    \label{loss}
\end{figure}
\begin{figure}
    \centering
    \includegraphics[width=0.4\textwidth]{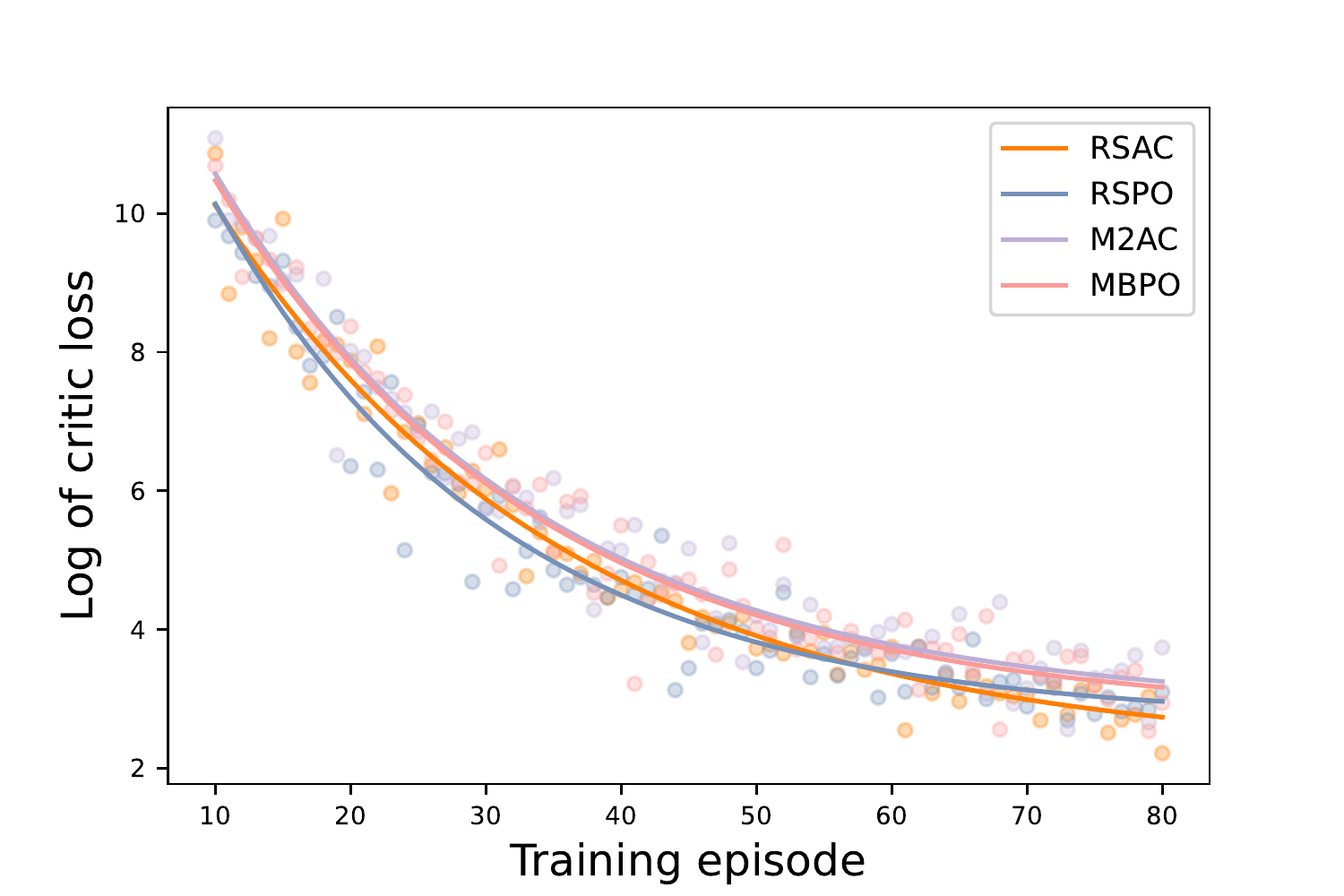}
    \caption{Critic loss over training steps.}
    \label{log_loss}
\end{figure}

\noindent\textbf{Convergence of Critic Network:} To guarantee reliability of our algorithms, we also analyze the convergence of critic network. Figure~\ref{loss} illustrates the relationship between critic loss over training episodes, which appears that all the algorithms converge in less than 50 episodes. In addition, the decline pattern of the critic loss over first 30 training episodes also shows that RSAC (orange) and RSPO (dark blue) converge faster than pure MBPO (purple) and M2AC (pink) at early stage, indicating that RSRS has helped our proposed method to be relatively efficient in terms of computational cost. The smaller box inside Figure~\ref{loss} shows the convergence in detail. In addition, Figure~\ref{log_loss} depicts the logarithm of critic loss value over training episodes and we can see that critic loss of RSAC (orange) and RSPO (dark blue) have consistently lower error compared with MBPO and RSPO without RSRS.


\begin{figure}
    \centering
    \includegraphics[width=0.4\textwidth]{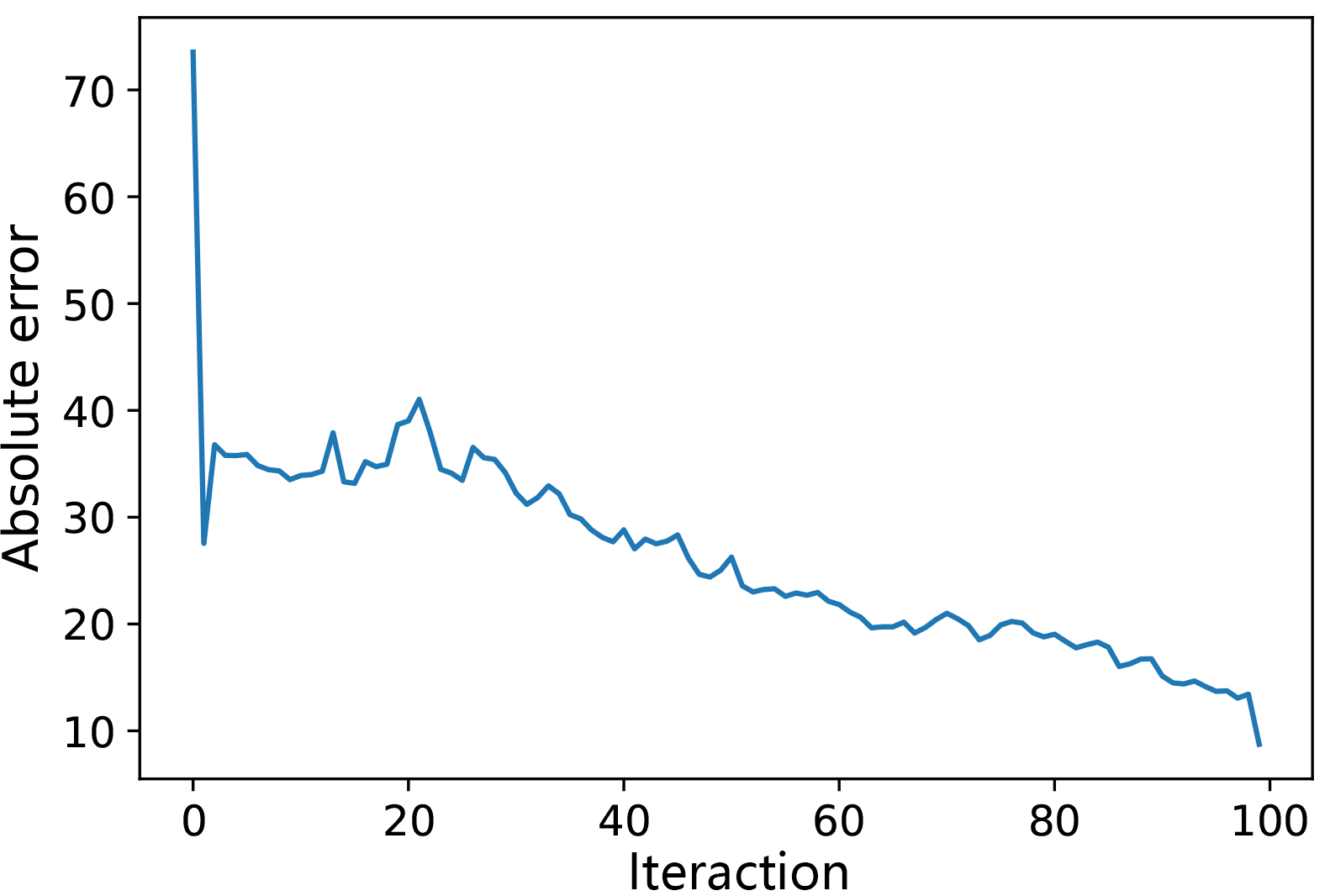}
    \caption{Absolute error of cumulative shares}
    \label{predicted_shares_error}
\end{figure}

\begin{figure}
    \centering
    \includegraphics[width=0.4\textwidth]{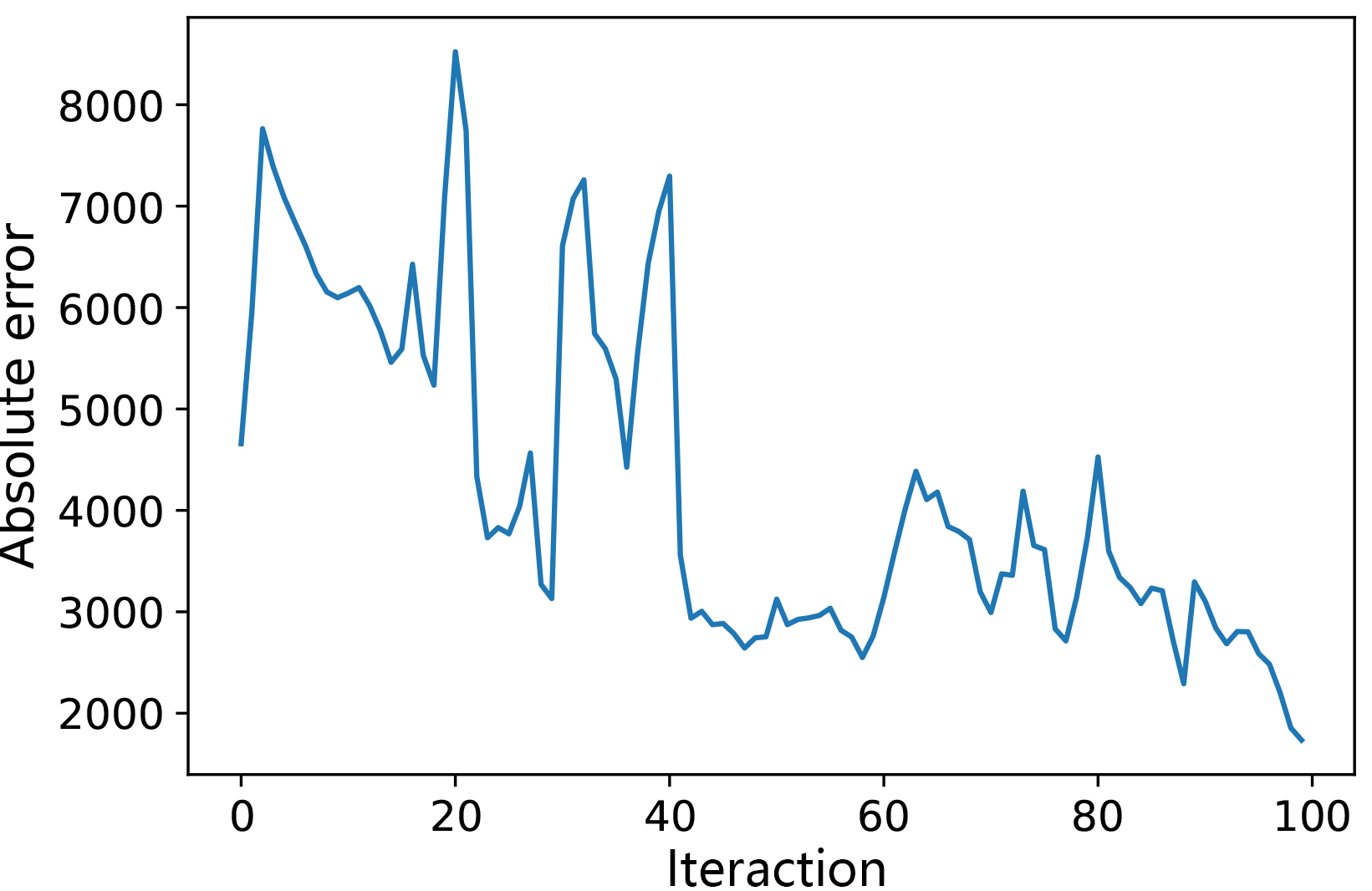}
    \caption{Absolute error of balance}
    \label{predicted_balance_error}
\end{figure}

\noindent\textbf{Analysis of Transition Dynamic Model:} The RL algorithms involved in this article are all model-based, so we also analyze the performance of our dynamic models. When training the transitional dynamics with quadruple data $(s, a, s', r)$ obtained from real environment, for any given $(s, a)$, we can predict $s'$ in virtual environment and therefore we also check the convergence of variable $s'$. Since our state is high dimensional, here we take some components of state 's' as an illustration. For example, Figure~\ref{predicted_shares_error} depicts the absolute error between predicted 'next state' and the ground truth of 'next state' projected onto the coordinate 'cumulative shares' (denoted as $W_t$ in Table~\ref{table_1}). It is shown that with the training iterations increasing by, the predict shares from transition dynamics getting closer to the true shares given by real environment. Moreover, Figure~\ref{predicted_balance_error} shows the error projected onto the component 'balance' which is the cash in the account over time (denoted as $B_t$ in Table 1). 

\section{Conclusion}
In this paper, we propose a novel strategy which brings resistance support and relative strength (RSRS) indicator into some powerful model-based RL methods for stock trading optimization. 

The RSAC and RSPO strategies take advantage of both the timing selection nature from technical analysis in finance market and the intelligent agent-environment interaction properties of model-based RL algorithms, leading to improvements in terms of stability, risk management and profit gain, especially the resistance to big marketing crisis. Experiment results based on 30 stocks selected form S$\&$P 500 market show that the proposed RSAC and RSPO strategies can effectively strengthen the performance of the model-based algorithms. For example, during COVID-19 pandemic period, our strategies highly improve the performance of maximum drawdown and annualized return measurements. To guarantee our observed results to be reliable, we also check the convergence of agent’s critic loss and prediction error of the dynamic model. Therefore, it inspires us that combining techniques from finance and statistics with RL algorithms is an interesting research direction to study.

However, there are still some challenges that are open to us. For example, as we known that there is close relationship between RL and stochastic optimal control. After setting up this question as a stochastic optimal control problem, how to prove the solution’s uniqueness and properties in appropriate functional space, is still unknown to us. Moreover, the analysis of generalization error bound of model-based RL is also important for us to design more adaptive and stable RL algorithms.


\bibliographystyle{unsrt} 
\bibliography{main}
\end{document}